\documentclass[conference]{IEEEtran}
\usepackage{fancyhdr}
\usepackage[numbers,sort&compress]{natbib}
\usepackage[bookmarks=true,breaklinks=true,colorlinks,citecolor=blue,linkcolor=blue,urlcolor=blue]{hyperref}
\usepackage{amsmath,amssymb,amsfonts}
\usepackage{amsthm}
\usepackage{algorithm, algorithmic}
\usepackage{graphicx}
\usepackage{textcomp}
\usepackage{xcolor}
\usepackage{threeparttable}
\usepackage{multirow}
\usepackage{makecell}
\usepackage{pifont}

\theoremstyle{definition}

\newtheorem{definition}{Definition}

\newtheorem{property}{Property}

\def\BibTeX{\rm B\kern-.05em{\sc i\kern-.025em b}\kern-.08em
    T\kern-.1667em\lower.7ex\hbox{E}\kern-.125emX}

\renewcommand\fbox{\fcolorbox{blue}{white}}

\title{Switch-Less Dragonfly on Wafers: A Scalable Interconnection Architecture based on \\ Wafer-Scale Integration}
\author{\IEEEauthorblockN{Yinxiao Feng and Kaisheng Ma\IEEEauthorrefmark{1}}
\IEEEauthorblockA{Institute for Interdisciplinary Information Sciences (IIIS) \\ 
Tsinghua University, 
Beijing, China}
}

\begin{document}
\maketitle
\pagestyle{plain}

 \thispagestyle{fancy}
 \lhead{}
 \rhead{}
 \chead{}
 \lfoot{\footnotesize{
 SC24, November 17-22, 2024, Atlanta, Georgia, USA
 \newline 979-8-3503-5291-7/24/\$31.00 \copyright 2024 IEEE}}
 \rfoot{}
 \cfoot{}
 \renewcommand{\headrulewidth}{0pt}
 \renewcommand{\footrulewidth}{0pt}

\begingroup\renewcommand\thefootnote{*}
\footnotetext{Corresponding author. Email: kaisheng@mail.tsinghua.edu.cn}
\endgroup


\begin{abstract}
  Existing high-performance computing (HPC) interconnection architectures are based on high-radix switches, which limits the injection/local performance and introduces latency/energy/cost overhead. The new wafer-scale packaging and high-speed wireline technologies provide high-density, low-latency, and high-bandwidth connectivity, thus promising to support direct-connected high-radix interconnection architecture.

  In this paper, we propose a wafer-based interconnection architecture called \textit{Switch-Less-Dragonfly-on-Wafers}. By utilizing distributed high-bandwidth networks-on-chip-on-wafer, costly high-radix switches of the \textit{Dragonfly} topology are eliminated while increasing the injection/local throughput and maintaining the global throughput. Based on the proposed architecture, we also introduce baseline and improved deadlock-free minimal/non-minimal routing algorithms with only one additional virtual channel. Extensive evaluations show that the \textit{Switch-Less-Dragonfly-on-Wafers} outperforms the traditional switch-based \textit{Dragonfly} in both cost and performance. Similar approaches can be applied to other switch-based direct topologies, thus promising to power future large-scale supercomputers.
\end{abstract}


\section{Introduction}
Mainstream high-performance computing (HPC) interconnection architectures are based on switches/routers. High-radix IO modules and switches enable very low-diameter network topologies, \textit{e.g.}, 2 switch-to-switch hops for Slim Fly~\cite{Besta_SlimFlyCost_2014} and PolarFly~\cite{Lakhotia_PolarFlyCostEffectiveFlexible_2022}, 3 hops for Dragonfly~\cite{Kim_TechnologyDrivenHighlyScalableDragonfly_2008}, and 4 hops for three-stage Fat-Tree~\cite{Stunkel_HighspeedNetworksSummit_2020}. However, high-radix switches are limited in the port number and bandwidth per link. 400G/800G is the maximum bandwidth provided by current Ethernet or InfiniBand adapters/switches~\cite{Routray_NewFrontiers800G_2020,Minkenberg_CoPackagedDatacenter_2021,_NVIDIAMQM9700NS2FQuantum_}. The limited physical channels connecting endpoints to the switch significantly constrain the local performance (injection bandwidth), which is critical for some workloads such as AI~\cite{Hoefler_HammingMeshNetworkTopology_2022}. Besides, high-radix switches are expensive and introduce additional latency and energy overhead~\cite{Barroso_DatacenterComputerDesigning_2019,Popa_CostComparisonDatacenter_2010,Greenberg_CostCloudResearch_2008,Popoola_EnergyConsumptionSwitchcentric_2018}. On the other hand, modern computing chips by themselves can provide abundant IO and switching bandwidth no
weaker than a regular switching chip~\cite{Ignjatovic_WormholeAITraining_2022,Elster_NvidiaHopperGPU_2022,Fischer_D17nmML_2023}, thus introducing the motivation to fully utilize the local bandwidth of computing chips~\cite{Hoefler_HammingMeshNetworkTopology_2022}.

In recent years, a new advanced packaging technology called \textit{wafer-scale-integration} promises to densely integrate tens of chips and provide ultra-high on/off-wafer bandwidth~\cite{DouglasYu_TSMCPackagingTechnologies_2021,Chun_InFO_SoWSystemonWaferHigh_2020}. For example, a tile of DOJO achieves 10TB/s on-wafer bisection bandwidth and 36 TB/s off-wafer aggregate bandwidth~\cite{Chang_DOJOSuperComputeSystem_2022}, which is far beyond any existing switch. Therefore, if the chips can be directly interconnected with high-bandwidth and low-latency, it not only improves the network performance but also promises to avoid using costly high-radix switches. However, scaling wafer-scale systems out for large-scale supercomputers still faces many challenges. \textbf{1)} Existing wafer-based systems, including \textit{Waferscale Processor}~\cite{Pal_Designing2048Chiplet14336Core_2021}, \textit{Wafer-Scale GPU}~\cite{Pal_ArchitectingWaferscaleProcessors_2019}, \textit{Wafer-Scale Engine (WSE)}~\cite{Cerebras_WaferScaleDeepLearning_2019, Lauterbach_PathSuccessfulWaferScale_2021, Lie_CerebrasArchitectureDeep_2022}, and \textit{DOJO}~\cite{Chang_DOJOSuperComputeSystem_2022, Talpes_DOJOMicroarchitectureTesla_2022, GaneshVenkataramanan_ComputeEnablingAI_2022, Talpes_MicroarchitectureDOJOTesla_2023}, are based on the 2D-mesh topology, which is not scalable due to the large diameter. \textbf{2)} The off-wafer bandwidth has a significant gap with the on-wafer bandwidth, which places higher demands on the hierarchy and configurability. \textbf{3)} Besides, interconnecting 2D-mesh-on-wafer by high-radix topologies introduces serious routing problems. The on-chip and off-chip routing must be designed and evaluated jointly rather than separately.

Motivated by these, we propose a new interconnection architecture called \textit{\textbf{Switch-less Dragonfly on Wafers}}. By utilizing distributed high-bandwidth networks-on-chip-on-wafer, we build a scalable wafer-based Dragonfly network without high-radix switches. The critical issues, including scalability, throughput, diameter, latency, energy, and cost, are quantitatively analyzed and discussed. We also give a simple minimal/non-minimal routing algorithm and a method to reduce the virtual-channel number. Extensive evaluations, including physical layout and cycle-accurate simulations on various workloads, are conducted based on the architecture. The contributions of this paper can be summarized as follows:
\begin{itemize}
  \item We propose a switch-less method to build the Dragonfly topology. Costly high-radix switches are eliminated while improving injection/local throughput and maintaining global throughput.
  \item The wafer-based interconnection architecture is a whole new frontier. We scale out existing 2D-mesh-on-wafer to large-scale high-radix network-of-wafers, achieving much better scalability than any existing wafer-based network.
  \item We introduce a simple baseline minimal/non-minimal routing algorithm, and novel labeling and interconnection methods are used to reduce the VC number. Only one additional virtual channel against traditional Dragonfly is required to achieve deadlock-free routing in the switch-less Dragonfly.
  \item Similar approaches can be applied to other switch-based direct topologies, including but not limited to Slim Fly~\cite{Besta_SlimFlyCost_2014}, PolarFly~\cite{Lakhotia_PolarFlyCostEffectiveFlexible_2022}, and HyperX~\cite{Ahn_HyperXTopologyRouting_2009}.
\end{itemize}

\section{Background \& Motivation}

\subsection{Wafer-Scale Integration}
\label{sec:wafer-scale}
\begin{figure}[htb]
  \centering
  \includegraphics[width=0.98\linewidth]{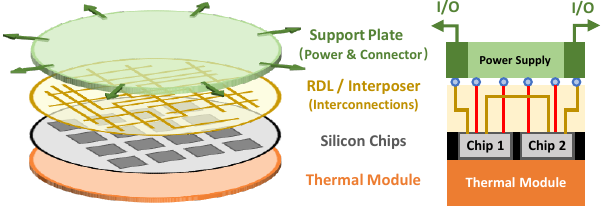}
  \caption{Profile of the InFO-SoW integration technology. Connectors and power modules are solder-joined to the InFO wafer~\cite{Chun_InFO_SoWSystemonWaferHigh_2020}. \label{fig:profile}}
\end{figure}
\subsubsection{Technology Introduction}
The traditional chip is implemented on a monolithic die, whose area is limited by the lithographic reticle (\textit{e.g.} $26mm \times 33mm$ for ASML lithography~\cite{_MaskReticleWikiChip_, Huang_WaferLevelSystem_2021}). Advanced packaging technologies integrate multiple chiplets within a package thus breaking through the ``Area Wall''. As shown in Fig.~\ref{fig:profile}, by using \textit{Integrated-Fan-Out-System-on-Wafer (InFO-SoW)} technology~\cite{DouglasYu_TSMCPackagingTechnologies_2021, Chun_InFO_SoWSystemonWaferHigh_2020}, tens of known-good chiplets, as well as power and thermal modules, are integrated into a whole wafer (diameter $300mm$). Compared with the traditional system, the wafer-scale integration eliminates using substrates and PCBs while achieving higher integration/interconnection density and energy efficiency.


\subsubsection{Wafer-based interconnection} In the past few years, many fantastic wafer-scale systems have emerged. The Tesla \textit{DOJO} integrates 25 D1 dies with an area of 645~$mm^2$~\cite{Fischer_D17nmML_2023}, resulting in a total silicon area exceeding 16,000~$mm^2$~\cite{Talpes_MicroarchitectureDOJOTesla_2023}. The \textit{WSE-2} designed by \textit{Cerebras} uses field stitching and achieves 850,000 cores (2.6 trillion transistors) on a wafer~\cite{Lie_CerebrasArchitectureDeep_2022}. All existing systems adopt 2D-mesh as the on-wafer topology because it is implementation-friendly and scheduling-friendly. However, planar topologies are insufficient to scale out. For example, the \textit{DOJO} supercomputer scales out the system by a larger 2D-mesh of wafers, resulting in a large diameter of up to 30 wafer-to-wafer hops~\cite{Chang_DOJOSuperComputeSystem_2022}. To reduce the diameter, a centralized switch is used to connect all the edges of the enormous 2D-mesh, which leads to limited scalability and a fault-tolerance problem~\cite{Talpes_DOJOMicroarchitectureTesla_2022}.

\subsection{HPC Network Fabric}
Almost all current HPC network architectures are based on switches. However, high-radix switches are very costly. A switch with $10 \times$ the bisection bandwidth often costs about $100 \times$ more~\cite{Barroso_DatacenterComputerDesigning_2019}. An InfiniBand switch with 64 400G ports is priced over \$40,000~\cite{_NVIDIAMQM9700NS2FQuantum_}.
The latency and power consumption of high-performance switches cannot be ignored either. The port-to-port latency of an InfiniBand switch is up to 200ns~\cite{Katebzadeh_EvaluationInfiniBandSwitch_2020}, and the power consumption of the switch can be up to 1.7 $KW$~\cite{_NVIDIAMQM9700NS2FQuantum_}. Meanwhile, the single physical link limits the injection bandwidth and local bandwidth between two terminals. For example, two servers are connected to a $64$-port 400G switch, whose total switching bandwidth is 25.6Tb/s, but the communication bandwidth between the two servers is only 400Gb/s.


\begin{table}[tbh]
  \fontsize{8pt}{10pt}\selectfont
  \centering
  \setlength{\tabcolsep}{1pt}
  \renewcommand\arraystretch{1.1}
  \caption{External communication and switching capability \\ of several datacenter chips \label{table:spec}}
  \begin{tabular}{|c|c|c|c|c|c|c|}
    \hline
    \textbf{Category}           & \multicolumn{3}{c|}{\textbf{Switching Chip}} & \multicolumn{3}{c|}{\textbf{Computing Chip}}                          \\
    \hline
    \textbf{Specification}      & \makecell{\textbf{NVSwitch}\vspace{-2pt}                                                                             \\ \cite{Ishii_NvlinkNetworkSwitchNvidia_2022} } & \makecell{\textbf{Tofino2}\vspace{-2pt} \\ \cite{Agrawal_IntelTofino212_2020} }      & \makecell{\textbf{Rosetta}\vspace{-2pt} \\ \cite{DeSensi_InDepthAnalysisSlingshot_2020}}               & \makecell{\textbf{H100} \vspace{-2pt} \\ \cite{Elster_NvidiaHopperGPU_2022,Choquette_NVIDIAHopperH100_2023}} & \makecell{\textbf{EPYC} \vspace{-2pt} \\ \cite{Naffziger_AMDChipletArchitecture_2020, Troester_AMDNextGeneration_2023}} & \makecell{\textbf{DOJO\,D1}\vspace{-2pt} \\ \cite{Fischer_D17nmML_2023}} \\
    \hline
    \textbf{Physical Lanes}     & 128                                          & 256                                          & 256  & 36  & 128 & 576 \\
    \hline
    \textbf{Data-rate}\,(Gbps)  & 100                                          & 50                                           & 50   & 100 & 32  & 112 \\
    \hline
    \textbf{Throughput}\,(Tb/s) & 12.8                                         & 12.8                                         & 12.8 & 3.6 & 4   & 63  \\
    \hline
  \end{tabular}
\end{table}

In recent years, with advances in high-speed wireline and packaging technologies, computing chips have become more powerful in NoC and IO throughput. As shown in TABLE~\ref{table:spec}, the NVIDIA H100 chip has 36 lanes of 100G link (3.6Tb/s IO bandwidth in total)~\cite{Choquette_NVIDIAHopperH100_2023,Ishii_NvlinkNetworkSwitchNvidia_2022}, and the Tesla \textit{DOJO} D1 chip has 576 lanes of 112G-SerDes (63Tb/s IO bandwidth in total)\cite{Fischer_D17nmML_2023}. The total external bandwidth and NoC throughput of current high-end computing chips are already at the same level as mainstream switching chips and even exceed some high-end switches. Therefore, many interconnection networks, including TofuD\cite{Ajima_TofuInterconnect_2018}, \textit{TPU}~\cite{Jouppi_TPUV4Optically_2023}, \textit{Wormhole}~\cite{Ignjatovic_WormholeAITraining_2022}, and \textit{DOJO}~\cite{Talpes_MicroarchitectureDOJOTesla_2023}, are using local interfaces and on-chip networks to scale out through direct (but low-radix) topologies. The injection/local bandwidth of these networks can be much higher than the limited bandwidth through a switch.

\subsection{State-of-the-Art Interconnection Networks}
\begin{figure}[tbh]
  \centering
  \includegraphics[width=0.98\linewidth]{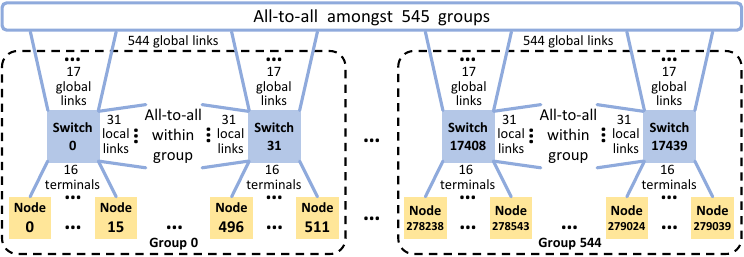}
  \caption{The Dragonfly-based Slingshot topology. Switches are fully connected within groups, and groups are also all-to-all connected. \label{fig:slingshot}}
\end{figure}
\subsubsection{Dragonfly} Three supercomputers of Top5, Frontier (\#1), Aurora (\#2), and LUMI (\#5), all adopt the Slingshot interconnect~\cite{_November2023TOP500_}. As shown in Fig.~\ref{fig:slingshot}, the \textit{Dragonfly} is the default topology for Slingshot~\cite{Kim_TechnologyDrivenHighlyScalableDragonfly_2008, DeSensi_InDepthAnalysisSlingshot_2020}. Several switches are fully connected between each other, forming a group, and multiple groups are also all-to-all connected.


\subsubsection{Diameter 2 Topologies} Slim Fly~\cite{Besta_SlimFlyCost_2014} and PolarFly~\cite{Lakhotia_PolarFlyCostEffectiveFlexible_2022} are two topologies towards Moore bound. PolarFly leverages silicon-photonic co-package~\cite{Maniotis_ScalingHPCNetworks_2020,Minkenberg_CoPackagedDatacenter_2021} to achieve more than 96\% of the theoretical peak with cost-effectiveness, which is a good example of innovating interconnection architecture through new technologies.

\begin{figure*}[tbh]
  \centering
  \includegraphics[width=0.99\linewidth]{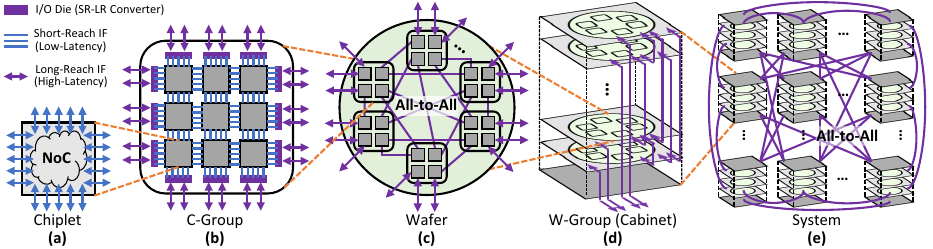}
  \caption{Hierarchical architecture of the wafer-based switch-less Dragonfly. (a) A chiplet has an on-chip network and several short-reach low-latency interfaces used for interconnection. (b) Several chiplets are connected by a planar topology (2D-mesh as the default), forming a C-group. The remaining short-reach interfaces at the edges of the C-group are converted to long-reach interfaces for upper-level high-radix interconnection. (c)(d) Each wafer consists of several C-groups, and several wafers form a W-group. All C-groups in a W-group are fully-connected. (e) All the w-groups in the system are also fully-connected, just as the Dragonfly topology.\label{figure:architecture}}
\end{figure*}

\subsubsection{HammingMesh} People have noted that the local bandwidth of existing switch-based networks is under-provisioned while current high-end chips have abundant IO and switching capability~\cite{Hoefler_HammingMeshNetworkTopology_2022}, which is also the major motivation of this paper. Using local 2D-mesh networks and a global Fat-Tree, the \textit{HammingMesh} provides high local bandwidth at a low cost with high scheduling flexibility.

\section{Architecture}
The following symbols are used in the description:

\renewcommand\arraystretch{1.1}
\begin{tabular}{lp{7cm}}
  $n$ & the number of interfaces (IO ports) of a chiplet                          \\
  $m$ & the scale of the 2D-mesh of chiplets in a C-group                         \\
  $k$ & the number of external interfaces of a C-group                            \\
  $a$ & the number of C-groups in a wafer                                         \\
  $b$ & the number of wafers in a W-group                                         \\
  $h$ & the number of global ports of a C-group used to connect to other W-groups \\
  $g$ & the number of W-groups in the system                                      \\
  $N$ & the total number of terminals/endpoints/chiplets                          \\
\end{tabular}

\subsection{Topology Description}
\label{sec:topology}
As shown in Fig.~\ref{figure:architecture}, the wafer-based switch-less Dragonfly architecture consists of 5 physical levels: chiplet, C-group, wafer, w-group, and system. Compared with the traditional switch-based Dragonfly~\cite{Kim_TechnologyDrivenHighlyScalableDragonfly_2008}, the chiplet is equivalent to the terminal (processor), the C-group is equivalent to the Dragonfly switch (router), and the W-group is equivalent to the Dragonfly router group.

\subsubsection{Chiplet} As shown in Fig.~\ref{figure:architecture}(a), the chiplet is the smallest component of the system. Each chiplet has an on-chip network and $n$ interconnection interfaces. The total IO ports, including memory and other peripherals, can be much more, but we focus only on the interconnection interfaces. These physical links are originally short-reach (\textit{e.g.}, UCIe~\cite{_UniversalChipletInterconnect_2023} or XSR SerDes~\cite{_CommonElectricalCEI_2022}) but have low latency and power consumption.

\subsubsection{C-Group} Chiplets are clustered into a chiplet-group by an on-wafer planar network as shown in Fig.~\ref{figure:architecture}(b). We adopt 2D-mesh as the default topology in the C-group because it is shortly-connected and implementation-friendly. A C-group consists of $m \times m$ chiplets. If each chiplet has $n/4$ ports at each edge, then a C-group has a total of $k=nm$ peripheral external ports. A C-group is equivalent to a switch in the traditional Dragonfly topology, with switching functionality realized through on-chip and intra-C-group interconnections. All the $k$ short-reach (SR) external interfaces of the C-group are converted to long-reach (LR) interfaces (\textit{e.g.}, LR SerDes~\cite{_CommonElectricalCEI_2022} and optic~\cite{Maniotis_ScalingHPCNetworks_2020}) through conversion modules to support the high-radix connectivity of the upper level.

\subsubsection{Wafer \& W-Group} As shown in Fig.~\ref{figure:architecture}(c)(d), each wafer consists of $a$ C-groups, and each W-group consists of $b$ wafers.
All $ab$ C-groups in a W-group are fully connected: each C-group connects to every other $a-1$ C-groups on the same wafer and every other $a(b-1)$ C-groups on the other wafers. It is feasible for $a=1$, then a whole wafer is a C-group, and there is no on-wafer all-to-all interconnection. When $a>1$, due to the wiring distance limitation, the logical on-wafer all-to-all connections are implemented off-wafer physically, which is further illustrated in Sec.~\ref{sec:on-wafer-long-distance}. The W-group is equivalent to the group with $ab$ switches in the traditional Dragonfly topology. Due to the ultra-high density of wafer-scale integration, one cabinet can hold an entire group that occupies a dozen cabinets in the traditional datacenter.

\subsubsection{System} The entire system has $g$ W-groups. As shown in Fig.~\ref{figure:architecture}(e), all W-groups are also fully connected: each connects to the other $g-1$  W-group by at least one link. Subtracting $ab-1$ interfaces used for local intra-W-group connections, the maximum number of global ports of a C-group is $h = k - ab + 1$, and the total number of W-groups in the system is $g = abh + 1$.

\subsection{Analysis}
\subsubsection{Scalability}
The total number of terminals (chiplets) in the wafer-based switch-less Dragonfly network described in Sec.~\ref{sec:topology} is:
\begin{equation}
  \label{eq:scale}
  N = abm^2 \times g = abm^2[ab(mn-ab+1)+1].
\end{equation}
Using a very small configuration $(a,b,m,n) = (2,4,2,6)$, the total chiplet number can reach 1K. The scale of the traditional Dragonfly network is bounded by the switch radix. However, in the switch-less Dragonfly, the functionality of the switch is realized by the network-of-chiplet in the C-group; therefore, the network scale can be very huge. Nevertheless, the scalability of the switch-less Dragonfly is constrained by two main factors:
\begin{itemize}
  \item \textbf{The physical scale of the wafer.} The maximum number of terminals (chiplets) that can be integrated within a C-group is limited by the area of the wafer (diameter $300mm$). With current technologies, a wafer can fit more than 64 server chips~\cite{SkyJuice_MonolithicSapphireRapids_2022}, which is a considerable scale.
  \item \textbf{The performance of the chiplet network within the C-group.} Forwarding through the network is not as straightforward as forwarding through a non-blocking switch. Therefore, as the scale increases, the intra-C-group network may become the bottleneck due to the competition of the intra/inter-C/W-group traffic. Related issues are further discussed in the following subsections.
\end{itemize}

\subsubsection{Throughput}
\label{sec:throughput}
If the bandwidth of all physical links is $1$ flit/cycle, the global saturation throughput (injection rate) $T_\text{global}$ of the switch-less Dragonfly can be estimated by the bisection bandwidth $B_C$ and the topology~\cite{Dally_PrinciplesPracticesInterconnection_2004}:
\begin{equation}
  \label{eq:global}
  \begin{aligned}
    T_\text{global} & < \frac{2B_C}{N} = \frac{(g/2)^2 \times 2 \times 2}{N}             \\
                    & = \frac{(mn-ab+1)}{m^2} \ [\text{flits}/\text{cycle}/\text{chip}].
  \end{aligned}
\end{equation}
For the traditional Dragonfly, the global-local ratio $h/t\approx 1/2$ maintains load-balance because each packet traverses one global and two local channels~\cite{Kim_TechnologyDrivenHighlyScalableDragonfly_2008}. In the switch-less Dragonfly, the global-local ratio can also be adjusted to about $1/2$ when $ab \approx (2/3)k = (2/3)mn$, $m^2 \approx (1/2)ab$. In this case, the theoretical global throughput limit in Equation (\ref{eq:global}) reaches $1$ flit/cycle/chip, the same as the traditional Dragonfly. Therefore, a reasonable configuration to achieve both globally load-balance and high-throughput is:
\begin{equation}
  \label{eq:configuration}
  \left\{
  \begin{array}{l}
    n = 3m, \\
    ab = 2m^2,
  \end{array}
  \right.
\end{equation}
As for local throughput, the injection rate in the switch-based Dragonfly is bounded by the single physical link between the chip and the switch ($1$ flit/cycle/chip). In the switch-less Dragonfly, chiplets in the C-group are connected through a network with multiple physical links, thus can achieve higher local throughput. The local intra-W-group saturation injection rate $T_\text{local}$ can be estimated as Equation (\ref{eq:local}):
\begin{equation}
  \label{eq:local}
  T_\text{local}  < \frac{(ab/2)^2\times 2 \times 2}{abm^2} = \frac{ab}{m^2} = 2 \ [\text{flits}/\text{cycle}/\text{chip}],
\end{equation}
twice as much as the throughput of the switch-based Dragonfly with the configuration of Equation (\ref{eq:configuration}). Since 2D-mesh is adopted in the C-group, the theoretical intra-C-group saturation throughput $T_\text{cg}$ can be estimated as Equation (\ref{eq:cg}):
\begin{equation}
  \label{eq:cg}
  T_\text{cg} < \frac{(nm/4)\times 2 \times 2}{m^2} = \frac{n}{m} = 3 \ [\text{flits}/\text{cycle}/\text{chip}],
\end{equation}
which is also much better than the traditional switch-based Dragonfly. Therefore, the wafer-based switch-less Dragonfly can achieve higher injection/local throughput than the traditional switch-based Dragonfly. \textbf{However, bottlenecks can still exist due to the competition for the intra-C-group bandwidth and the imbalance of traffic distribution.} The total full-duplex bisection bandwidth $B_\text{cg}$ of the 2D-mesh-in-C-group is
\begin{equation}
  B_\text{cg} = \frac{nm}{2} = \frac{k}{2} \ [\text{flits}/\text{cycle}],
\end{equation}
which is half of the $k$-port non-blocking switch ($k$ flits/cycle).
As a result, the inter-C-group traffic will compete with the intra-C-group traffic for the bandwidth provided by 2D-mesh. Therefore, to prevent the intra-C-group network from becoming the bottleneck under extreme traffic, a larger intra-C-group link bandwidth or higher-bandwidth topology, such as HexaMesh~\cite{Iff_HexaMeshScalingHundreds_2023}, is required. Higher intra-C-group bandwidth is easy and affordable to achieve by wafer-level integration. For example, the UCIe die-to-die interface can provide $1317$ GB/s/$mm$ die edge density ($947$ GB/s/$mm^2$ area density) on the wafer~\cite{_UniversalChipletInterconnect_2023}, much larger than traditional off-chip links.

\subsubsection{Diameter}
The diameter of the Dragonfly network consists of one global hop and two local hops. Therefore, in the worst case, a packet in the switch-less Dragonfly goes through four C-groups: source C-group, destination C-group, and two intermediate C-groups. Each 2D-mesh-based C-groups has a diameter of $2(m-1)$ chiplet-to-chiplet hops. At the same time, each inter-C-group hop requires two additional SR-LR conversion hops. Therefore, the diameter (only off-chip hops are counted) of the wafer-based switch-less Dragonfly can be described as Equation (\ref{eq:diameter}):
\begin{equation}
  \label{eq:diameter}
  D = \underbrace{H_g + 2H_l}_\text{Dragonfly hops} + \underbrace{(8m-2)H_{sr}}_\text{intra-C-group hops} ,
\end{equation}
where $H_g$ is a global hop, $H_l$ is a local hop, $H_{sr}$ is an on-wafer short-reach hop or a SR-LR hop. For comparison, the diameter of the traditional switch-based Dragonfly is $H_g + 2H_l + 2H_l^*$, where $H_l^*$ is a hop from the terminal (processor) to the switch, whose typical cost is similar to a local hop. The rough cost of these hops is compared in TABLE~\ref{tab:hop}.
\begin{table}[tbh]
  \centering
  \setlength{\tabcolsep}{3pt}
  \renewcommand\arraystretch{1.1}
  \caption{Comparison of hop cost~\cite{DeSensi_InDepthAnalysisSlingshot_2020, Sella_FECKilledCutThrough_2018,_CommonElectricalCEI_2022, Katebzadeh_EvaluationInfiniBandSwitch_2020,  Synopsys_DesignWareDietoDie112G_2021, Frankel_ProspectsOpticalTransceivers_2021, DavideTonietto_EnergyEfficiencySerial_2023,Navaridas_UnderstandingImpactArbitration_2024,Feng_HeterogeneousDietoDieInterfaces_2023} \label{tab:hop}}
  \begin{tabular}{|c|c|c|c|c|}
    \hline
                             & $H_g$              & $H_l$              & $H_{sr}$ & $H_\text{on-chip}$ \\
    \hline
    \textbf{Physical Medium} & Optical Cable      & Copper Cable       & RDL      & Metal Layer        \\
    \hline
    \textbf{Latency} (ns)    & $150 + \text{ToF}$ & $150 + \text{ToF}$ & $\sim 5$ & $\sim 1$           \\
    \hline
    \textbf{Energy} (pj/bit) & $20 + $            & $20 + $            & $\sim 2$ & $\sim 0.1$         \\
    \hline
  \end{tabular}
\end{table}

\begin{table*}[ht]
  \fontsize{7.5pt}{9pt}\selectfont
  \centering
  \setlength{\tabcolsep}{2.5pt}
  \renewcommand\arraystretch{1.1}
  \caption{Comparison of key specifications between the switch-less Dragonfly and other topologies \label{table:comparison}}
  \begin{tabular}{|c|c|c|c|c|c|c|c|c|c|}
    \hline
    \textbf{Interconnection Network}     & \textbf{Chip-radix} & \textbf{SW-radix}   & \textbf{\#\,Switch} & \textbf{\#\,Cabinet} & \textbf{\#\,Processor} & \textbf{Cable Number\,/\,Length}    & $\bf T_\textbf{local}$ & $\bf T_\textbf{global}$ & \textbf{Diameter}                               \\
    \hline
    \textbf{2D-Mesh\,\&\,Switch (DOJO)}  & $8$                 & $60$                & $1$                 & $2$                  & $450$                  & $/$                                 & $1.6$                  & $0.53$                  & $2H_l^* + 18H_{sr}$                             \\
    \hline
    \textbf{Three-Stage Fat-Tree}        & $1$                 & \multirow{3}*{$64$} & $5120$              & $608$                & $65536$                & $N=197K$                            & $1$                    & $1$                     & \multirow{3}*{$2H_g + 2H_l + 2H_l^*$}           \\
    \textbf{Three-Stage Fat-Tree}        & 4                   &                     & $20480$             & $896$                & $65536$                & $N=786K$                            & 4                      & $4$                     &                                                 \\
    \textbf{Three-Stage F-T (3:1 Taper)} & 4                   &                     & $14336$             & $960$                & $98304$                & $N=655K$                            & 4                      & $4/3$                   &                                                 \\
    \hline
    \textbf{1-Plane Hx4Mesh}             & $4$                 & \multirow{2}*{$64$} & $5120$              & $352$                & \multirow{2}*{$65536$} & $N=197K$                            & $2$                    & $1/2$                   & \multirow{2}*{$2H_g + 2H_l + 2H_l^* + 4H_{sr}$} \\
    \textbf{4-Plane Hx4Mesh}             & $16$                &                     & $20480$             & $640$                &                        & $N = 786K$                          & $8$                    & $2$                     &                                                 \\
    \hline
    \textbf{Co-Packaged PolarFly (p=32)} & $1$                 & $64$                & $4033$              & $504$                & $129056$               & $N=129K$                            & $1$                    & $1$                     & $2H_g + 2H_{sr}$                                \\
    \hline
    \textbf{Dragonfly (Slingshot)}       & $1$                 & $64$                & $17440$             & $2180$               & $279040$               & $N$=$698K \,/\, L$=$154K$$\cdot$$E$ & $1 (1)$                    & $1$                     & $H_g + 2H_l + 2H_l^*$                           \\
    \hline
    \textbf{Switch-less Dragonfly}       & $12$                & $/$                 & $0$                 & $545$                & $279040$               & $N$=$419K \,/\, L$=$73K$$\cdot$$E$  & $3 (2)$                    & $1$                     & $H_g + 2H_l + 30H_{sr}$                         \\
    \hline
  \end{tabular}
\end{table*}

Ignoring protocol layers and considering only the physical layer, the latency of a short-reach hop generally comes from the PHY (\textit{e.g.} UCIe and XSR SerDes~\cite{_CommonElectricalCEI_2022}). When the transmission distance exceeds $100 mm$, forward error correction (FEC) must be introduced, significantly increasing the latency by tens of nanoseconds~\cite{Sella_FECKilledCutThrough_2018}. Above $10 m$, electro-optical (E-O) conversion is necessary, and time-of-flight (ToF) in fiber can no longer be ignored. For instance, the latency of a $10 m$ optical link can easily be up to $200 ns$, which is approximately $40 \times$ higher than the on-wafer short-reach link. Besides the latency, the energy cost of long-distance hops is also much larger than the on-wafer hops. In the traditional Dragonfly, each packet must traverse these two local hops; however, in the switch-less Dragonfly, the number of short-reach hops is not always high.

\subsubsection{Collective Communication}
\begin{figure}[tb]
  \centering
  \includegraphics[width=0.9\linewidth]{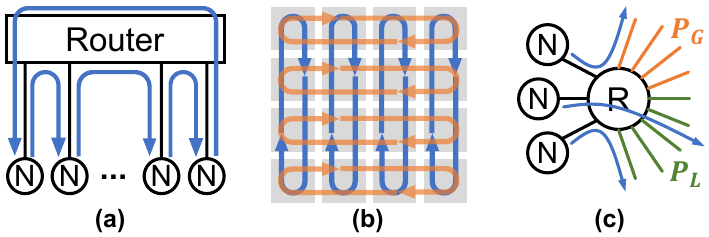}
  \caption{Bottleneck of the switch-less Dragonfly in collective communication. \label{fig:allreduce} (a) Ring AllReduce algorithm; (b) 2D algorithm for AllReduce within the 2D-mesh-based C-group; (c) Local/global link underutilization due to injection bandwidth limit.}
\end{figure}
The throughput analysis in Sec.~\ref{sec:throughput} is based on the assumption that the traffic is uniformly distributed across the bisection links. Under real workloads, the bottleneck of the switch can be more visible. As shown in Fig.~\ref{fig:allreduce}, if the ring-based AllReduce algorithm is performed on a switch-based topology, the maximum bandwidth of the ring is $1$ flit/chip/cycle, and the latency of $N$ nodes is $O(N)$. On the 2D-mesh, as shown in Fig.~\ref{fig:allreduce}(b), 2D algorithms can be performed to reduce the latency to $O(\sqrt{N})$~\cite{Kumar_HighlyAvailableData_2020, Luczynski_NearOptimalWaferScaleReduce_2024,Sensi_SwingShortcuttingRings_2024}. Besides, \textit{bidirectional pipelined rings} can also be used to further reduce the latency~\cite{Hoefler_HammingMeshNetworkTopology_2022}. For inter-router communication, the injection bandwidth can also become the bottleneck. As shown in Fig.~\ref{fig:allreduce}(c), in a typical Dragonfly, terminals take up only a quarter of the switch ports (bandwidth). As a result, it is hard for a collective algorithm to fully utilize all the bandwidth, especially for small-scale jobs or hierarchical algorithms~\cite{Feng_OptimizedMPICollective_2022}. For the 2D-mesh-based C-group, the injection bandwidth is adequate thus the total off-C-group bandwidth can be fully utilized. 

\subsection{Comparison by Case Study}
\label{sec:casestudy}
We compare the specifications of several typical HPC interconnection networks under specific configurations in TABLE~\ref{table:comparison}. All links are assumed to have the same bandwidth (normalized as 1), and $T_\text{local}$ is the theoretical throughput of a subset of processors (\text{e.g}, a group of the Dragonfly and a Hx4Mesh board of HammingMesh). All the topologies attempt to fully utilize the 64-port switch. We use a switch-less Dragonfly of the same scale as the Slingshot shown in Fig.~\ref{fig:slingshot} for comparison~\cite{DeSensi_InDepthAnalysisSlingshot_2020}. The configuration of the switch-less Dragonfly is as follows:
\begin{itemize}
  \item $n=12, m=4$, Every chiplet has $3$ external ports at each edge, and chiplets form the C-group by a 4$\times$4 2D-mesh.
  \item $a=4, b=8$, Each wafer has 4 C-groups (64 chiplets), and eight wafers form a W-group (512 chiplets).
  \item Each W-group has a total of 544 off-W-group ports, so there are up to $g=545$ W-groups and a total of $N=279040$ chiplets.
\end{itemize}


\subsubsection{Bandwidth Trade-off} The injection bandwidth can become the bottleneck for most existing switch-based topologies, including Fat-Tree, Dragonfly, and PolarFly. However, it is not easy to simply increase injection/ejection channels because available terminal ports are limited by the switch radix and network scale. Doubling the ports of a traditional endpoint results in doubling the requirement for the network building blocks. If we are willing to sacrifice the diameter and scalability, mesh/torus or DOJO-like topologies can provide adequate bandwidth for a small-scale system (hundreds of chips). Or, if we are willing to sacrifice the global throughput, the tapered Fat-Tree is a potential choice. Alternatively, the HammingMesh enables flexible configurations for different scales, diameters, bandwidth, and costs; however, it is still constrained by the Fat-Tree backbone. The \textit{switch-less Dragonfly on wafer} provides another approach to directly build high-radix networks without switches. The intra-C-group and intra-W-group local throughput reaches $3$ and $2$ flits/cycle/chip, respectively, which is much higher than the traditional switch-based networks. With high-bandwidth on-wafer interconnects, the throughput can be even higher; at the same time, the global throughput is maintained. In summary, we achieve high injection/local/global bandwidth, low diameter, low cost, and high scalability, simultaneously.


\subsubsection{PolarFly} The co-packaged PolarFly achieves the lowest diameter with integrated high-radix optical IO modules (OMs). PolarFly~\cite{Lakhotia_PolarFlyCostEffectiveFlexible_2022} does not discuss the in-package network in detail though it is critical for the overall performance. If there are multiple processors and OMs in each package, besides all external IO ports, additional processor-to-OM and OM-OM ports inside the package are required. These intra-package hops are regarded as short-reach hops, equivalent to on-wafer hops. With current technologies, it is hard to integrate $32$ high-performance processors and multiple centralized high-radix IO modules in a single package. However, with a wafer-scale integration and a similar switch-less approach, the \textit{switch-less PolarFly on wafer} promises to provide a more scalable and cost-effective solution.

\subsubsection{Cost} \label{sec:cost} The switch-less Dragonfly avoids using costly high-radix switches, thus significantly reducing the overall cost, including switches themselves and related power/cooling infrastructure. With wafer-scale integration, substrates and PCBs are also eliminated while providing affordable high-bandwidth interconnects. $1 mm^2$ silicon-on-wafer ($<1$\$) provides more than $800$ GB/s~\cite{_UniversalChipletInterconnect_2023} on-wafer bandwidth, much cheaper than the traditional inter-rack IOs and cables. Besides, wafer-scale integration also increases the density, thus reducing the physical size of the entire system. According to ~\cite{_HewlettPackardEnterprise_}, one cabinet can host 64 blades, each consisting of 2 nodes; therefore, assuming 8 switches are at the top-of-rack (ToR), the Slingshot system requires 2180 cabinets in total. Besides, we also assume 32 core switches (except the ToR switch) can be placed in a cabinet for Fat-Tree-based networks. Short-reach 2D-mesh-on-PCB and co-package can increase the density, thus each cabinet is supposed to host 16 Hx4Mesh boards or 8 PolarFly co-packages (twice chips per cabinet). Conservative estimation suggests that the density of a single cabinet can increase by at least $4 \times$ through wafer-scale integration~\cite{Tesla_TeslaAIDay_2022, Chang_DOJOSuperComputeSystem_2022,Lauterbach_PathSuccessfulWaferScale_2021}. As a result, the wafer-based switch-less Dragonfly only requires 545 cabinets (8 wafers per cabinet) to hold a system as large as the maximum Slingshot. If the Slingshot is flatly laid out in the datacenter at scale $E \times E$, the total cable length of inter-cabinet links can be estimated by cabinet-to-cabinet distance at $154K\cdot E$. For comparison, the local cable of switch-less Dragonfly is very short (intra-cabinet), and the total cable length is only $73K\cdot E$, less than half of the switch-based Dragonfly. Besides, all the terminal adapters and cables are also eliminated. In summary, the benefits of wafer-level integration and switch-less are all-encompassing, saving numerous datacenter building blocks.

\subsection{Architecture Variations}
\label{sec:varision}
\subsubsection{Small-Scale Networks} HPC systems are not always very large. A single-chiplet C-group with only 12 external ports can be used to build a system of up to $333$ chips (nodes). In this case, short-reach interfaces and conversion modules are not necessary. Besides, the inter-W-group interconnection can be eliminated; that is to say, the system is a single fully-connected W-group, whose diameter is only $H_l+(4m-2)H_{sr}$.


\subsubsection{Topology Variations} For many domain-specific workloads such as AI-training, the requirement for networks can be various~\cite{Hoefler_HammingMeshNetworkTopology_2022}. Therefore, the topology is supposed to be adjustable. First, the parameters ($a,b,m,n$) of the switch-less Dragonfly can be changed to achieve unbalanced local/global bandwidth.
Second, the topology of the intra-C-group network can be changed to HexaMesh~\cite{Iff_HexaMeshScalingHundreds_2023} or other topologies.
Third, C-groups within the W-group can be connected by a flatter topology (\textit{e.g.} 2D-flattened-butterfly), which consumes fewer local ports and is easier to lay out. Besides, other state-of-the-art topologies including but not limited to Slim Fly~\cite{Besta_SlimFlyCost_2014}, PolarFly~\cite{Lakhotia_PolarFlyCostEffectiveFlexible_2022}, and HyperX~\cite{Ahn_HyperXTopologyRouting_2009}, can also be built by integrating endpoints under a switch through a planar topology on the wafer.

\begin{figure}[t]
  \centering
  \renewcommand\fbox{\fcolorbox{red}{white}}
  \includegraphics[width=0.97\linewidth]{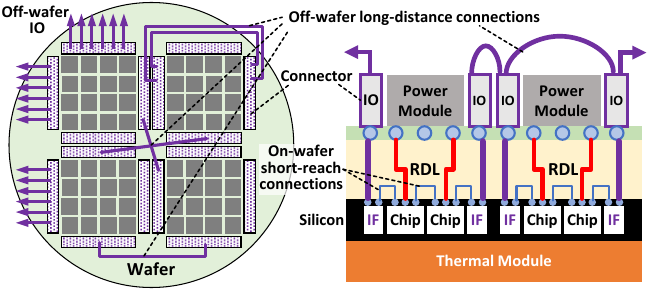}
  \caption{Wafer-level long-distance connectivity. All the edge IOs of each C-group are fanned out, and the long-distance wafer-level logical links are connected off-wafer physically. \label{fig:all-to-all-on-wafer}}
\end{figure}

\subsection{Wafer-Level Long-Distance Interconnection} 
\label{sec:on-wafer-long-distance}
As discussed above, when higher-radix topologies are used intra-C-group, or when there is more than one C-group on each wafer, wafer-level long-distance interconnections are required. However, due to the limitations of manufacturing, traditional technologies, such as field stitching~\cite{Lauterbach_PathSuccessfulWaferScale_2021,Flack_LithographicManufacturingTechniques_1992}, only allow short-distance wiring within a lithographic reticle. The advanced mask stitching technologies~\cite{Hou_WaferLevelIntegrationAdvanced_2017,Huang_WaferLevelSystem_2021,Hou_SupercarrierRedistributionLayers_2023} allows cross-reticle redistribution layer (RDL), and the reliability/quality of the wires across the stitching boundary is fine (negligible resistance contribution). However, though the stitched RDL promises to allow long-distance ($>100$ mm) wiring, the high-speed electrical signals may not be able to travel that far. Therefore, other technologies such as on-wafer repeaters~\cite{Han_BigChipChallenge_2023, Vaisband_CommunicationConsiderationsSilicon_2019, Chen_WaferscaleNetworkSwitches_2024} are necessary.

Nevertheless, the switch-less Dragonfly is still practical without any physical on-wafer long-distance wires. because the inter-C-group interconnections do not require high-density on-wafer wiring. For a wafer with 9 C-groups (smaller C-groups do not require wafer-scale integration), there are only 36 inter-C-group wafer-level channels, which can be implemented off-wafer by standard packaging and interconnections. As shown in Fig.~\ref{fig:all-to-all-on-wafer}, each C-group is manufactured as a single unit with high-density short-reach on-wafer wiring, but all edge IOs, no matter whether for on-wafer or off-wafer interconnection, are fanned out to off-wafer electrical/optical connectors~\cite{Wade_TeraPHYChipletTechnology_2020,Hsia_IntegratedOpticalInterconnect_2023,Kopp_SiliconPhotonicCircuits_2011}. Then, the long-distance wafer-level logical links are connected off-wafer physically by backplane or cables. For the system discussed in Sec.~\ref{sec:casestudy}, the total number of IO channels for a wafer is 192, and a practical layout of the C-group is presented in Fig.~\ref{fig:layout}.



\section{Interconnection and Routing Design}
Routing is one of the core problems of interconnection networks. In traditional switch-based Dragonfly, the minimal path is unique, and all ports of a switch are equivalent and directly connected; thus, the routing is simple. Kim \textit{et al.} achieved deadlock-free minimal routing by two virtual channels (VCs) and non-minimal routing by three VCs~\cite{Kim_TechnologyDrivenHighlyScalableDragonfly_2008}. However, in the switch-less Dragonfly, the switching functionality is realized by the distributed networks-on-chiplet; therefore, the ports of a C-group are non-equivalent, and channel dependencies among on-chip and off-chip networks can lead to potential deadlocks. Therefore, it is essential to illustrate the routing design of the entire network. In this section, we first introduce a simple baseline routing algorithm, and then present methods to reduce the number of virtual channels. Besides, the impact of intra-C-group networks is also discussed.

\begin{figure}[tbh]
  \centering
  \includegraphics[width=0.98\linewidth]{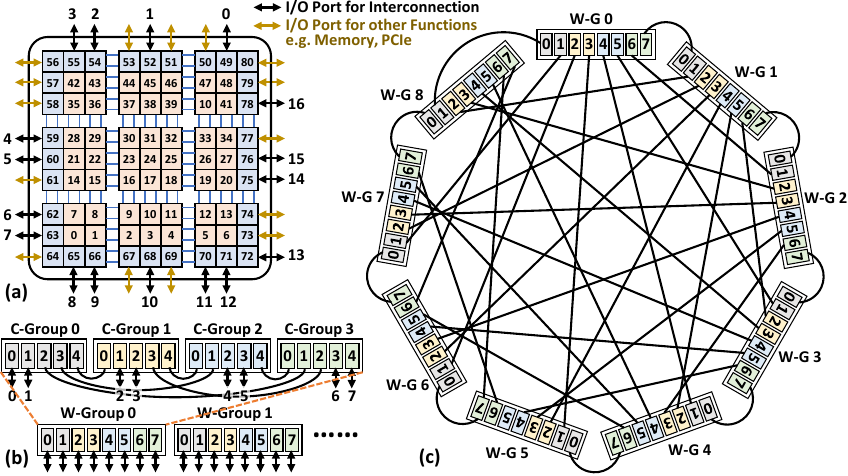}
  \caption{Intra/inter-C/W-group interconnection. (a) Each in-C-group node has a unique label; $k$ ports used for interconnection are also labeled. (b) C-groups are connected into multiple W-groups by local ports; The remaining ports are led out and re-labeled for global interconnection. (c) W-groups are fully connected. \label{fig:interconnection}}
\end{figure}

\subsection{Baseline Virtual-Channel-based Routing}
The interconnection is shown in Fig.~\ref{fig:interconnection}. In brief, the network is built in two steps: \textbf{1)} Label the port and fully connect C-groups into multiple W-groups. \textbf{2)} Relabel the remaining ports and fully connect all W-groups into a Dragonfly.

\begin{algorithm}[b]
  \begin{algorithmic}
    \REQUIRE Source: $(W_s, C_s, n_s)$, \\ Destination: $(W_d, C_d, n_d)$;
    \ENSURE Routing within C-group from node $n_i$ to $n_j$;
    \vspace{2pt}
    \STATE \textbf{Step 1:} {RWC($n_s$, $n_a$)}. $n_a \in C_s$ is the node that has the local channel to $C_b$, which has the global channel to $W_d$.
    \vspace{2pt}
    \STATE \textbf{Step 2:} Traverse the local channel from $n_a$ to $n_{b0} \in C_b$.
    \vspace{2pt}
    \STATE \textbf{Step 3:} {RWC($n_{b0}$, $n_{b1}$)}. $n_{b1} \in C_b$ is the node that has the global channel to $C_c \in W_d$.
    \vspace{2pt}
    \STATE \textbf{Step 4:} Traverse the global channel from $n_{b1}$ to $n_{c0} \in C_c$.
    \vspace{2pt}
    \STATE \textbf{Step 5:} {RWC($n_{c0}$, $n_{c1}$)}. $n_{c1} \in C_c$ is the node that has the local channel to $C_d \in W_d$.
    \vspace{2pt}
    \STATE \textbf{Step 6:} Traverse the local channel from $n_{c1}$ to $n_{d0} \in C_d$.
    \vspace{2pt}
    \STATE \textbf{Step 7:} {RWC($n_{d0}$, $n_{d}$)}.
  \end{algorithmic}
  \caption{\scshape Minimal routing in SW-less Dragonfly \label{alg:minimal}}
\end{algorithm}

As shown in Algorithm~\ref{alg:minimal}, the minimal routing algorithm in the switch-less Dragonfly from the source node $n_s$ of the source C-group $C_s$ of the source W-group $W_s$ to the destination node $n_d$ of the destination C-group $C_d$ of the destination W-group $W_d$ is accomplished in seven steps: three inter-C-group routing steps and four intra-C-group routing steps. The non-minimal routing is similar to the minimal routing but with two additional inter-C-group steps and two additional intra-C-group steps at an intermediate W-group. Deadlock-free routing within 2D-mesh-based C-group can simply follow existing algorithms (\textit{e.g.}, dimension-order and negative-first routing). Virtual channels (VCs) are used to avoid cross-C-group deadlocks in the switch-less Dragonfly. There are four kinds of situations for a minimal-routed packet in the C-group: source C-group $C_s$, intermediate C-group $C_b$, $C_c$, and destination C-group $C_d$. Therefore, we can simply use four VCs to avoid any cross-C-group deadlock by increasing the VC at each C-group. Similarly, six VCs can be used for deadlock-free non-minimal routing.

\subsection{VC Number Reduction}
When the VC number is limited, we also present methods to reduce the VC number. The basic idea is to achieve up*/down* deadlock-free routing~\cite{Schroeder_AutonetHighspeedSelfconfiguring_1991} in a larger subnetwork beyond a C-group.
If there is a valid up-first path for any source-destination pair within a W-group, the two VCs of the two C-groups can be merged into one VC. The up*/down* routing relies on proper labeling and interconnection. Definition~\ref{def:up-down} gives the type of all channels and ports. A feasible labeling method is stated in Property~\ref{prop:1}, which makes all ports consistently ordered and higher than the cores. The corresponding interconnection method is stated in Property~\ref{prop:2}, which organizes the different types of ports consistently from low to high: local ports to lower C-groups, global ports, and local ports to higher C-groups.

\begin{definition}
  A physical or virtual channel from node $(w_i, c_i, n_i)$ to node $(w_j, c_j, n_j)$ is \textit{up} if:
  \begin{itemize}
    \item $w_i < w_j$, \textbf{or}
    \item $w_i = w_j, c_i < c_j$, \textbf{or}
    \item $w_i = w_j, c_i = c_j, n_i < n_j$;
  \end{itemize}
  otherwise, the channel is \textit{down}. A port $P_s$ of a C-group or W-group is \textit{up} if the channel from $P_s$ to $P_d$ is \textit{up}; otherwise, the port is \textit{down}.
  \label{def:up-down}
\end{definition}

\begin{property}
  \label{prop:1}
  For the intra-C-group network,
  \begin{itemize}
    \item[$\bf c1.$] $\forall$ port-core pair $(p, n)$, $\exists$ a down-only path from $p$ to $n$ (\textit{i.e.} an up-only path from $n$ to $p$).
    \item[$\bf c2.$] $\forall$ port-port pair with label $(i, j), i < j$, $\exists$ an up-only path from $i$ to $j$ (\textit{i.e.} a down-only path from $j$ to $i$), \textbf{and}
  \end{itemize}
\end{property}

\begin{property}
  As shown in Fig.~\ref{fig:interconnection}(b), $\forall$ global port of the C-group, all \textit{down} local ports are at lower position, and all \textit{up} local ports are at higher position.
  \label{prop:2}
\end{property}

\begin{figure}[ht]
  \centering
  \includegraphics[width=0.98\linewidth]{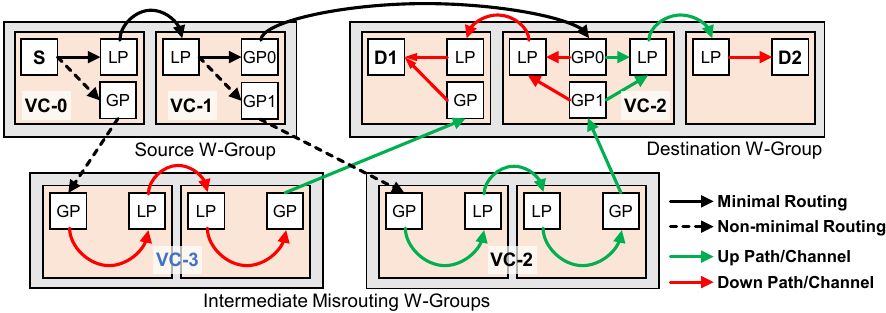}
  \caption{Minimal/non-minimal routing and virtual channel assignment in the switch-less Dragonfly. S is the source node, and D1/D2 are the destination nodes. \label{fig:routing}}
\end{figure}

As a result, any packet at the destination W-group has a valid up-first path to the destination: \textbf{1)} If the packet is at the core, it can reach the local port through an up-only path by Property~\ref{prop:1}($\bf c1$); and no matter the next local inter-C-group hop is \textit{up} or \textit{down}, it can then reach the destination core through a \textit{down-only} path. \textbf{2)} As shown in Fig.~\ref{fig:routing}, if the packet reaches the port node through a global channel, according to Property~\ref{prop:2} and Property~\ref{prop:1}($\bf c2$), there is a \textit{down-only} or \textit{up-only} path to the local port of the destination C-group; and then, according to Property~\ref{prop:1}($\bf c1$), there is a \textit{down-only} path to the destination core. Therefore, one VC can be reduced for minimal/non-minimal routing at the destination W-group.

Similarly, any packet that reaches the intermediate W-group by non-minimal routing has a consistent path from the entering global port to the leaving global port: According to Property~\ref{prop:2}, if the leaving C-group is higher than the entering C-group, the path is \textit{up-only}; otherwise, the path is \textit{down-only}. As shown in Fig.~\ref{fig:routing}, if we only allow non-minimal routing to a lower W-group from which there exists an \textit{up-only} path to the destination W-group, then the routing among the intermediate and destination W-group can be merged with unified up*/down* routing. If allowing non-minimal routing to other W-groups, one more VC is still required for the intermediate W-group.

In summary, the minimal routing requires three VCs: VC-0 and VC-1 for the source and intermediate C-groups of the source W-group, and VC-2 for the destination W-group. No more VC is required if only misrouting to a valid lower W-group; otherwise, one more VC-3 is required at the intermediate W-group.

\subsection{Intra-C-group Networks}
\label{sec:noc}
As stated in Property~\ref{prop:1}, two conditions for the intra-C-group network are required for up*/down* routing. Various intra-C-group network architectures can meet the conditions by trading off performance and complexity.

\begin{figure}[tb]
  \centering
  \includegraphics[width=0.98\linewidth]{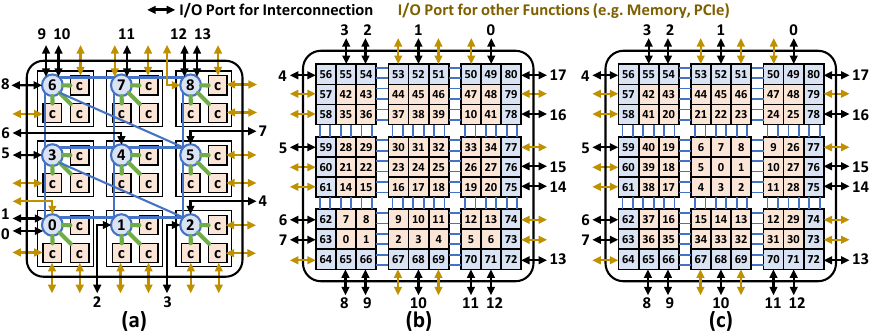}
  \caption{Network-in-C-group architectures and the labeling. (a) IO-router-based: all interconnection ports are connected to one on-chip router;  (b)(c) Mesh-based: interconnection ports are distributed at the edge of the NoC. \label{fig:noc}}
\end{figure}

The IO-router-based NoCs shown in Fig.~\ref{fig:noc}(a) are adopted by many chips, including the \textit{EPYC}~\cite{Troester_AMDNextGeneration_2023,Naffziger_AMDChipletArchitecture_2020}, TofuD~\cite{Ajima_TofuInterconnect_2018}, \textit{H100}~\cite{Choquette_NVIDIAHopperH100_2023}, and \textit{TPU}~\cite{Norrie_DesignProcessGoogle_2021}. The advantages of the IO-router-based NoCs are the isolation of on/off-chip traffic and the simplification of intra-C-group interconnection.  However, the IO router can become the bottleneck, and the chip-to-chip bandwidth does not scale with the chip scale. Fig.~\ref{fig:noc}(a) shows a valid intra-C-group interconnection and labeling method for IO-router-based chiplets by four physical channels.

The mesh-based NoCs can provide a more scalable injection bandwidth. Many recent multi-chip systems, including the \textit{Sapphire Rapids}~\cite{Nassif_SapphireRapidsNextGeneration_2022}, \textit{Wormhole}~\cite{Ignjatovic_WormholeAITraining_2022}, and \textit{DOJO}~\cite{Talpes_MicroarchitectureDOJOTesla_2023}, adopt such an architecture. Fig.~\ref{fig:noc}(b) shows a labeling method consistent with the on-chip routing, and Fig.~\ref{fig:noc}(c) shows another novel polar-system-based labeling method. Both two labeling methods meet the condition in Property~\ref{prop:1} but are different in design detail. For example, router-less rings can be implemented on the polar-system-labeled NoCs to reduce the complexity and detour~\cite{Liu_IMRHighPerformanceLowCost_2016,Alazemi_RouterlessNetworkonChip_2018}. A potential issue is the asymmetry of any such labeling method; however, since our labeling is software-based (the physical 2D-mesh is symmetric), it is possible to change the labeling method or mapping policy for different applications. More details are beyond the scope of this paper.

\section{Evaluation \& Discussion}
\begin{figure}[tbh]
  \centering
  \includegraphics[width=0.98\linewidth]{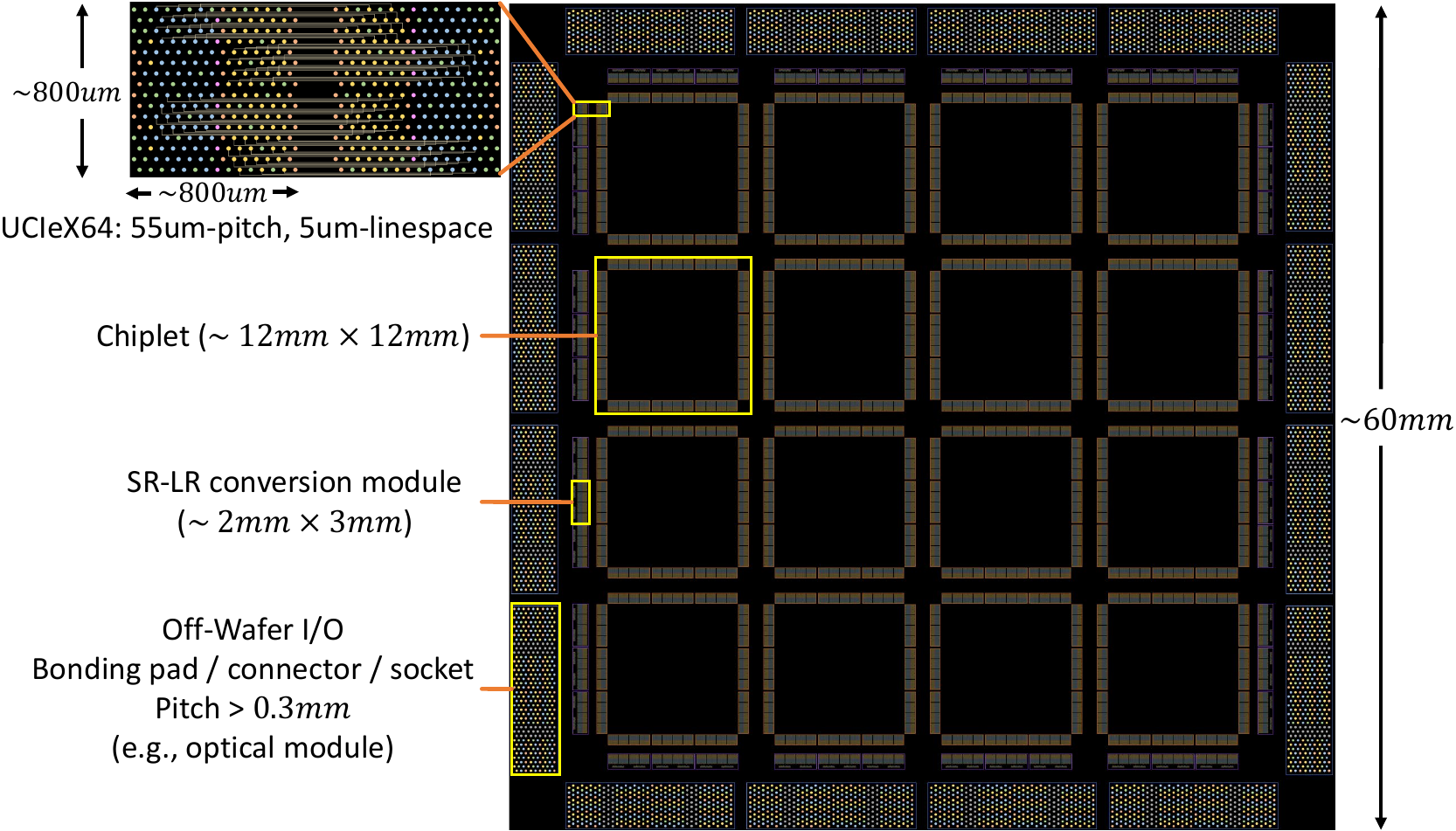}
  \caption{Layout of PHYs, chiplets, and IO connectors of a C-group.   \label{fig:layout}}
\end{figure}

\begin{figure*}[tb]
  \centering
  \includegraphics[width=0.99\linewidth]{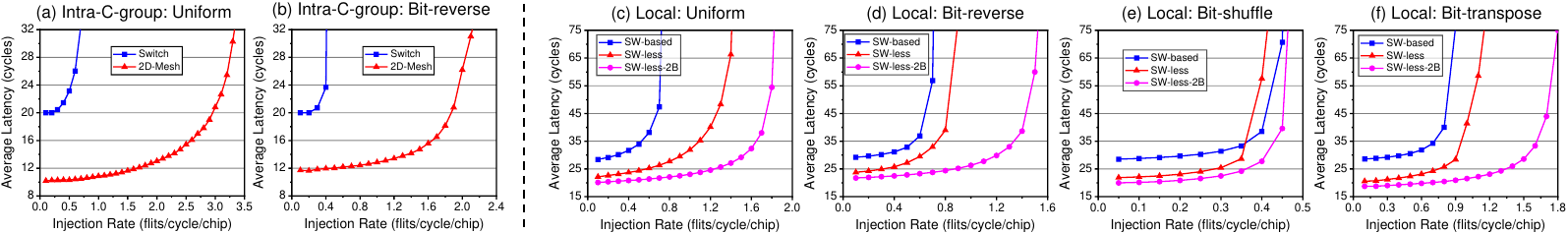}
  \caption{(a-b) Intra-C-group (intra-switch) and (c-f) local (intra-Dragonfly-group) performance under different traffic patterns. \label{figure:local-throughput}}
\end{figure*}
\subsection{Methodology}
\subsubsection{Layout} \label{sec:layout} To evaluate the feasibility of the implementation, we try to place and route a C-group on the wafer. The bump pitch and line space are assumed to be $55$um and $5$um on the wafer~\cite{DouglasYu_TSMCPackagingTechnologies_2021}. As shown in Fig.~\ref{fig:layout}, the layout includes placement of PHYs, chiplets, and IO connectors. Assuming the C-group consists of $16$ chiplets, each chiplet has $6$ physical channels at each edge. In our layout, $128$ lanes of UCIe (two $64\times$ PHY~\cite{_UniversalChipletInterconnect_2023}) are adopted at each on-wafer channel, achieving $4096$ Gb/s/port intra-C-group short-reach bandwidth. $8$ lanes of 112G SerDes (differential signal) are adopted at each off-C-group channel, achieving $896$ Gb/s/port long-reach bandwidth~\cite{Synopsys_DesignWareDietoDie112G_2021, _CommonElectricalCEI_2022}.
As a result, a C-group of $60mm\times 60mm$ size leads out 1536 pairs of differential ports ($\sim 5500$ IOs including the power and ground) in total. The total bisection and aggregation bandwidth of the on-wafer C-group is $12$TB/s and $20.9$TB/s, much larger than the highest-end switches. The layout also suggests that it is feasible to achieve multiples of bandwidth on-wafer with advanced packaging and interface technologies.
\begin{table}[ht]
  \centering
  \setlength{\tabcolsep}{5pt}
  \renewcommand\arraystretch{1.1}
  \caption{Default Parameters}
  \label{table:parameter}
  \begin{tabular}{|l|l|}
    \hline
    \textbf{Parameter}     & \textbf{Value}                 \\
    \hline
    Packet Length          & $4$ flits                      \\
    \hline
    Input Buffer Size      & $32$ flits                     \\
    \hline
    Base Link Bandwidth    & $1$ flit/cycle                 \\
    \hline
    Short-Reach Link Delay & $1$ cycle                      \\
    \hline
    Long-Reach Link Delay  & $8$ cycles                     \\
    \hline
    Simulation Time        & $10000$ cycles  \vspace{-1pt}  \\
                           & after $5000$ cycles warming up \\
    \hline
  \end{tabular}
\end{table}
\subsubsection{Simulator} CNSim~\cite{Feng_EvaluatingChipletbasedLargeScale_2024} is used to evaluate the performance. The default parameters used in simulations are shown in TABLE~\ref{table:parameter}. We do not set the long-reach link delay at the real value (hundreds of cycles), otherwise, the switch-less Dragonfly will always have a much lower latency due to the shorter diameter (3 \textit{v.s.} 5 hops).

\subsubsection{Workloads}
The evaluations use three kinds of network workloads: \textbf{(a) Unicast traffic patterns.} The \textit{uniform} and other permutation patterns~\cite{Dally_PrinciplesPracticesInterconnection_2004}, including bit-reverse, bit-shuffle, and bit-transpose, are evaluated. \textbf{(b) Adversarial traffic patterns.} We evaluate the \textit{hotspot} traffic pattern, which conducts communications within four of all W-groups, and the \textit{worst-case (WC)} traffic pattern, where each node in W-group $W_i$ sends traffic to a random node in W-group $W_{i+1}$~\cite{Kim_TechnologyDrivenHighlyScalableDragonfly_2008}. \textbf{(c) Collective traffic patterns.} We also evaluate the \textit{ring-based AllReduce} traffic pattern, where each chip (process) $i$ sends the $1/N$ segment to chip $(i+1)$ mod $N$ or sends two $1/2N$ segments to $(i-1)$ mod $N$ and $(i+1)$ mod $N$~\cite{Feng_OptimizedMPICollective_2022, Hoefler_HammingMeshNetworkTopology_2022}.

\subsubsection{Experiment Setup} The baseline is the standard switch-based Dragonfly. A switch's terminal, local, and global ports are configured at 4:7:5 for radix-16 and 8:15:9 for radix-32. As a result, the total (group, chip) numbers are (41, 1312) for radix-16 and (145, 18560) for radix-32. For the switch-less Dragonfly, local and global ports are configured as the same number but no terminal ports. All nodes in a C-group of the switch-less Dragonfly are connected by a 2D-mesh with low-latency on-wafer links. The links between C-groups and W-groups are configured the same as the switch-based Dragonfly. As discussed in Sec.~\ref{sec:throughput}, the 2D-mesh with uniform link bandwidth has limited bisection bandwidth compared with a non-blocking switch. Therefore, we also evaluate the configuration of higher intra-C-group bandwidth (labeled as ``2B/4B'' for $2\times$/$4\times$ intra-C-group bandwidth). It is also important to note that all the switches are modeled as single ideal high-radix routers; however, they are actually also implemented by distributed networks-on-chip~\cite{DeSensi_InDepthAnalysisSlingshot_2020,Ahn_ScalableHighradixRouter_2013}. The performance/energy overhead of the high-radix switches is underestimated in this paper.

\subsection{Performance}
\subsubsection{Local Throughput} Rather than connecting to the switch by a single physical channel, the switch-less Dragonfly adopts a 2D-mesh within the C-group. As a result, the theoretical local throughput of the switch-less Dragonfly is more than $1$ flit/cycle/chip. We evaluate the architecture by adopting a 2D-mesh of $2 \times 2$ chiplets with $2 \times 2$ on-chiplet network in the C-group ($4 \times 4$ on-chip routers in total). The C-group has $12$ external ports ($7$ for local and $5$ for global, equivalent to the radix-16 switch); therefore, each W-group has $8$ fully-connected C-groups (32 chips in total). As shown in Fig.~\ref{figure:local-throughput}(a), the saturation injection rate intra-C-group under uniform and bit-reverse traffic reaches $3.0$ and $2.0$ flits/cycle/chip, which is over $3 \times$ more than connecting to a switch. As for the intra-W-group throughput, although a traditional Dragonfly switch has $2 \times$ local ports than the terminal ports, the injection rate is still bounded by the single injection channel connecting to the switch. As shown in Fig.~\ref{figure:local-throughput}(c-f), except for the bit-shuffle pattern, the saturation injection rate intra-W-group can be $1.2-2\times$ larger. With double on-wafer bandwidth, the performance can be even better. However, the performance is not improved if the bottleneck is the inter-C-group links rather than the intra-C-group links (\textit{e.g.} bit-shuffle pattern shown in Fig.~\ref{figure:local-throughput}(e)). In summary, the switch-less Dragonfly achieves better local throughput without doubling the intra-C-group bandwidth.

\begin{figure}[tb]
  \centering
  \includegraphics[width=0.95\linewidth]{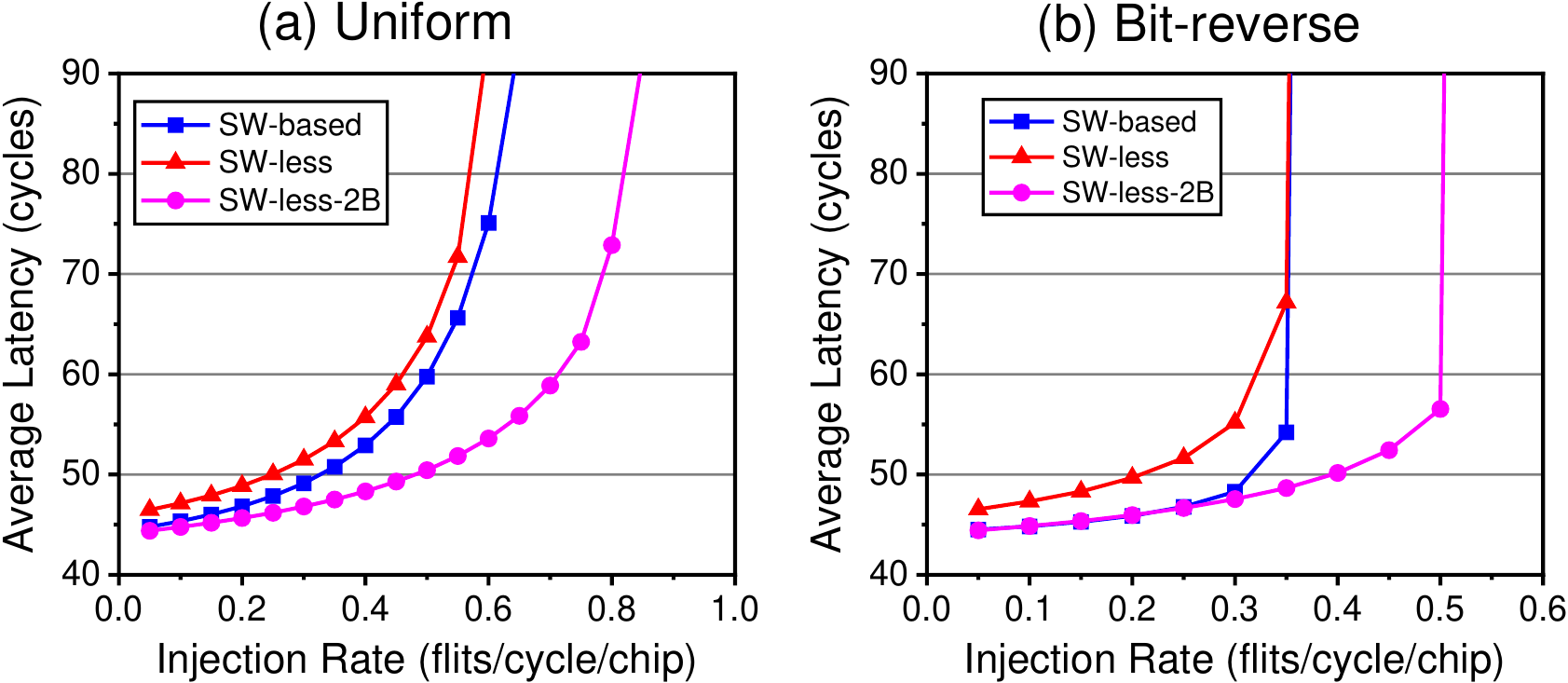}
  \caption{Global performance under the uniform and bit-reverse traffic patterns.   \label{figure:global-throughput}}
\end{figure}

\subsubsection{Global Throughput} We evaluate the global performance of the same radix-16 network. The whole network has 1312 chips (5248 on-chip nodes) in total. As shown in Fig.~\ref{figure:global-throughput}(a), if the intra-C-group link bandwidth is the same as the local/global link bandwidth, the overall performance under uniform traffic for the switch-less Dragonfly is slightly worse than the switch-based Dragonfly due to the limited bisection bandwidth of the 2D-mesh-in-C-group. If the intra-C-group link bandwidth is doubled, the bottleneck on the bisection bandwidth is eliminated; thus, the switch-less Dragonfly performs much better than the traditional Dragonfly. For the bit-reverse traffic pattern shown in Fig.~\ref{figure:global-throughput}(b), the result is similar. For small-scale networks, the switch-less Dragonfly maintains the global performance with uniform bandwidth and achieves better performance with higher intra-C-group bandwidth.

\begin{figure}[tb]
  \centering
  \includegraphics[width=0.95\linewidth]{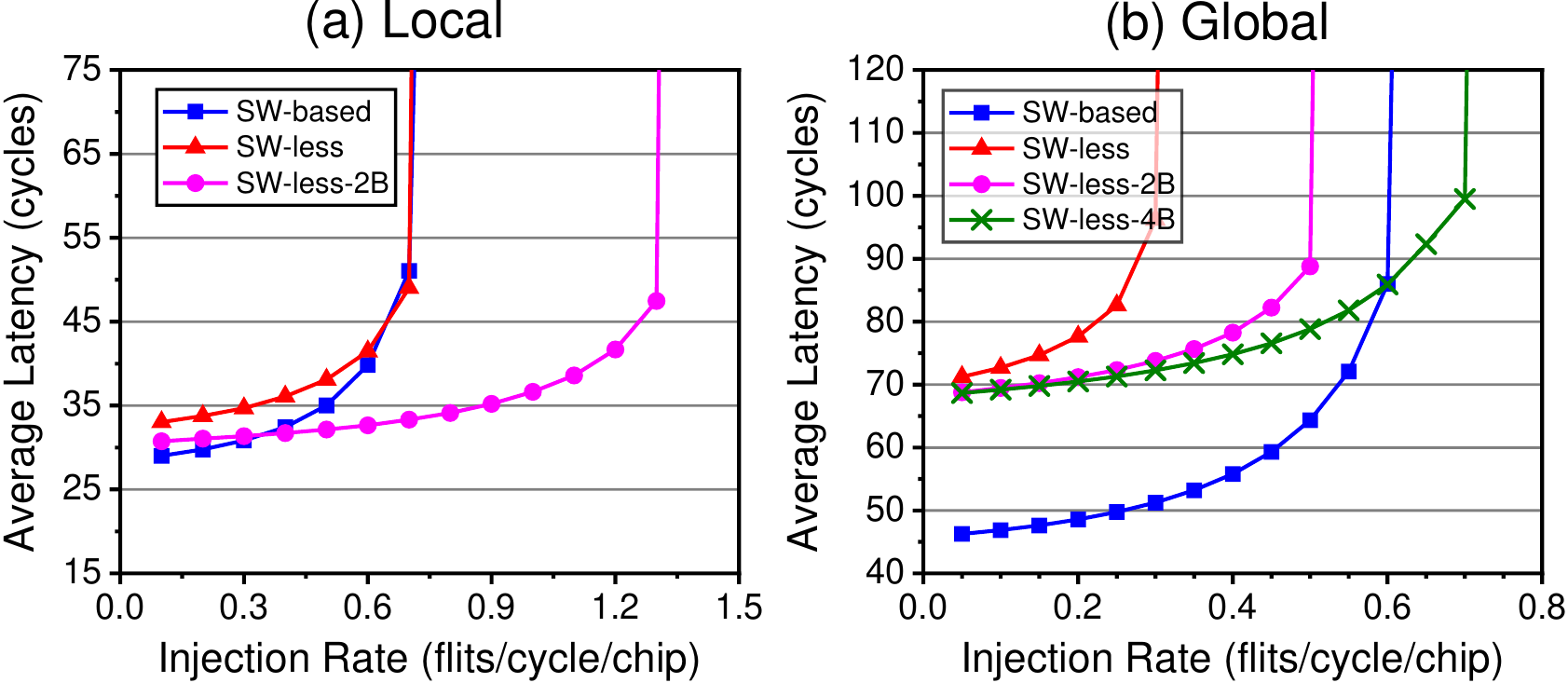}
  \caption{Performance scalability under the uniform traffic. \label{figure:large-scale}}
\end{figure}

\subsubsection{Scalability} We also evaluate the scalability of the switch-less Dragonfly by simulating large-scale networks. It is important to note that the absolute value of the latency of the switch-based Dragonfly is greatly underestimated for easier comparison. We build a large-scale system of 18560 chips (radix-32). As shown in Fig.~\ref{figure:large-scale}(a), the local performance of the large-scale switch-less Dragonfly is not as good as small-scale networks without doubling intra-C-group bandwidth. As shown in Fig.~\ref{figure:large-scale}(b), the global performance of the uniform-bandwidth switch-less Dragonfly is severely constrained by the limited bisection bandwidth of the 2D-mesh-in-C-group. That is intuitive and inevitable since we have eliminated thousands of powerful switches with non-blocking switching capability. Higher intra-C-group bandwidth is critical for removing the bottlenecks for extreme global traffic. As shown in Fig.~\ref{figure:large-scale}(b), the global throughput can be maintained or even improved after increasing the intra-C-group bandwidth. As we have analyzed and validated in Section~\ref{sec:throughput}, \ref{sec:cost}, and \ref{sec:layout}, it is feasible and affordable to achieve higher bandwidth on the wafer; or from another perspective, off-wafer bandwidth is reduced compared to the on-wafer bandwidth, just as the DOJO~\cite{Talpes_DOJOMicroarchitectureTesla_2022}.

\subsubsection{Misrouting} 
\begin{figure}[tb]
  \centering
  \includegraphics[width=0.98\linewidth]{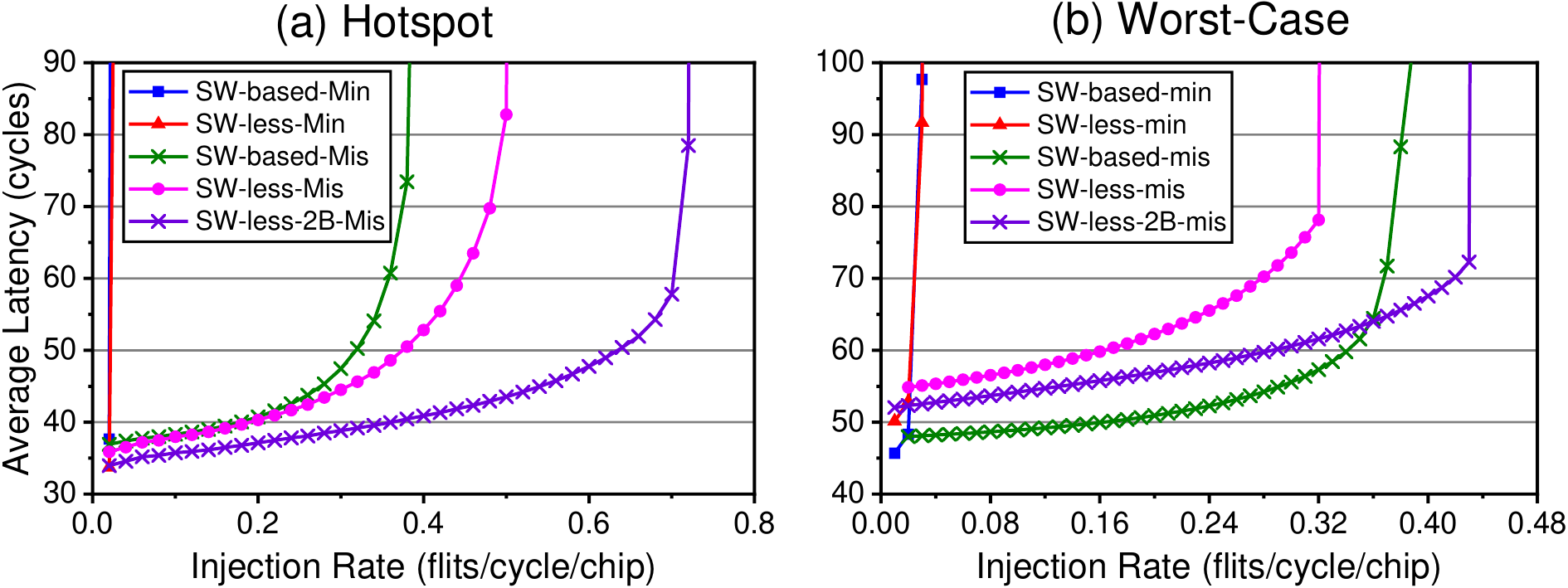}
  \caption{Performance of the minimal and non-minimal routing under the hotspot and wost-case traffic patterns.\label{figure:misrouting}}
\end{figure}

The minimal routing on the Dragonfly topology is insufficient for some unbalanced traffic; thus, non-minimal routing is required. We evaluate the non-minimal routing algorithm under the hotspot and worst-case traffic patterns. As shown in Fig.~\ref{figure:misrouting}, the performance by minimal routing is poor because only 3/40 global links are used for the hotspot traffic, and only 1/40 global links are used for the worst-case traffic. Therefore, distributing traffic to more global channels by non-minimal routing can reduce congestion. The simulation results show that the saturation injection rate by non-minimal routing is tens of times larger than the minimal routing. As shown in Fig.~\ref{figure:misrouting}(a), increasing the intra-C-group bandwidth can significantly improve the performance of the hotspot pattern because traffic congestion is also within the C-group. 

\begin{figure}[tb]
  \centering
  \includegraphics[width=0.95\linewidth]{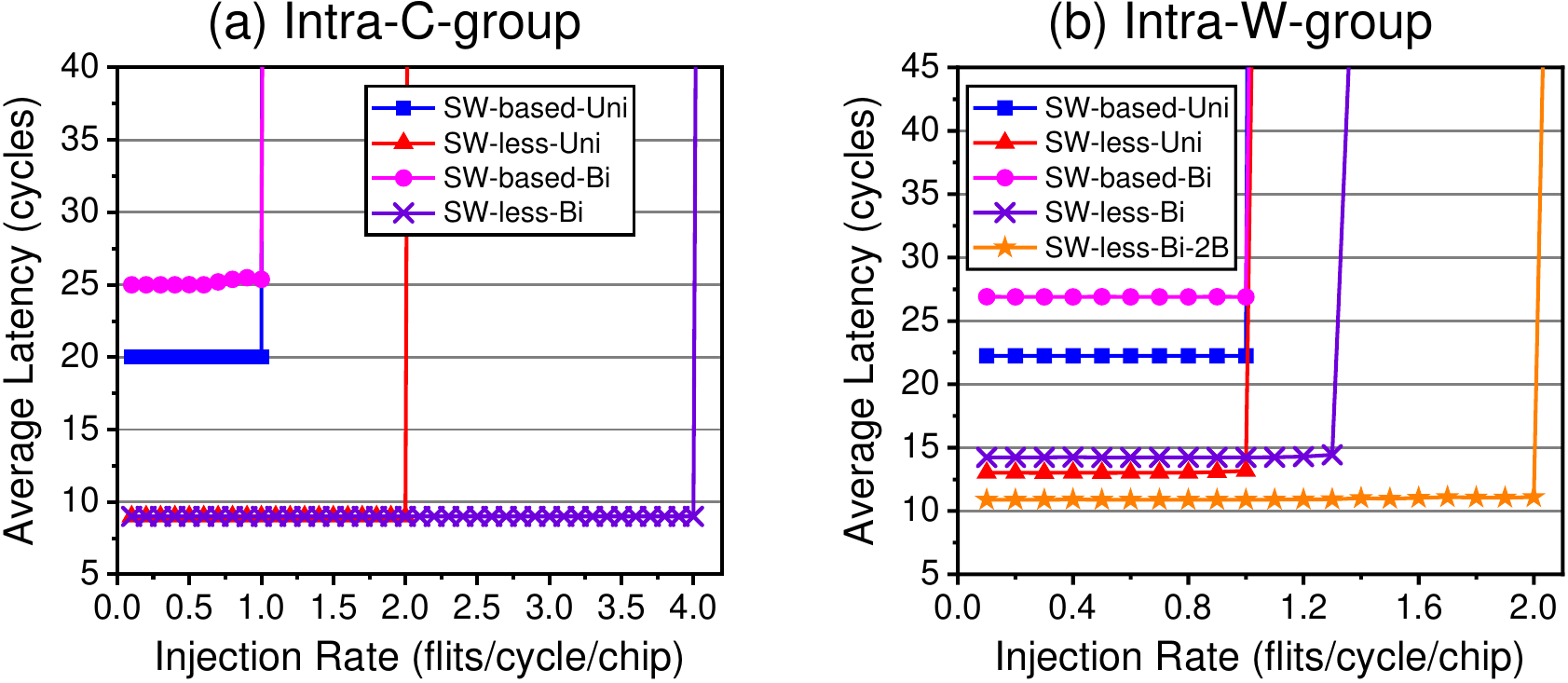}
  \caption{Performance of ring-based AllReduce algorithm within C-group and W-group. \label{fig:allreduce-throughput}}
\end{figure}

\subsubsection{AllReduce Traffic}
We also evaluated the AllReduce traffic based on the unidirectional and bidirectional rings. As shown in Fig.~\ref{fig:allreduce-throughput}(a), the saturation injection rate of the switch-based Dragonfly reaches $1$ flit/cycle/chip for intra-C-group AllReduce. The bidirectional ring does not improve the performance but introduces congestion at the ejection port and leads to higher latency. Meanwhile, since there are four injection/ejection ports per chip in the switch-less Dragonfly, the saturation throughput can reach $2$ and $4$ flits/cycle/chip through the unidirectional and bidirectional rings. If considering the on-wafer bandwidth can be multiple times more, the expected performance can be even higher. As shown in Fig.~\ref{fig:allreduce-throughput}(b), the performance of the intra-W-group AllReduce is bounded by the inter-C-group links. Without bidirectional rings, both the switch-based and switch-less Dragonfly reach the same throughput ($1$ flit/cycle/chip). With bidirectional rings, the switch-less Dragonfly can achieve a higher throughput of $1.3$ flits/cycle/chip, but still lower than the theoretical value due to the competition on the intra-C-group networks. By doubling the intra-C-group bandwidth, the intra-C-group bandwidth bottleneck is eliminated, thus the performance of inter-C-group AllReduce can reach $2$ flits/cycle/chip, twice that of the switch-based Dragonfly.


\begin{figure}[tb]
  \centering
  \includegraphics[width=0.98\linewidth]{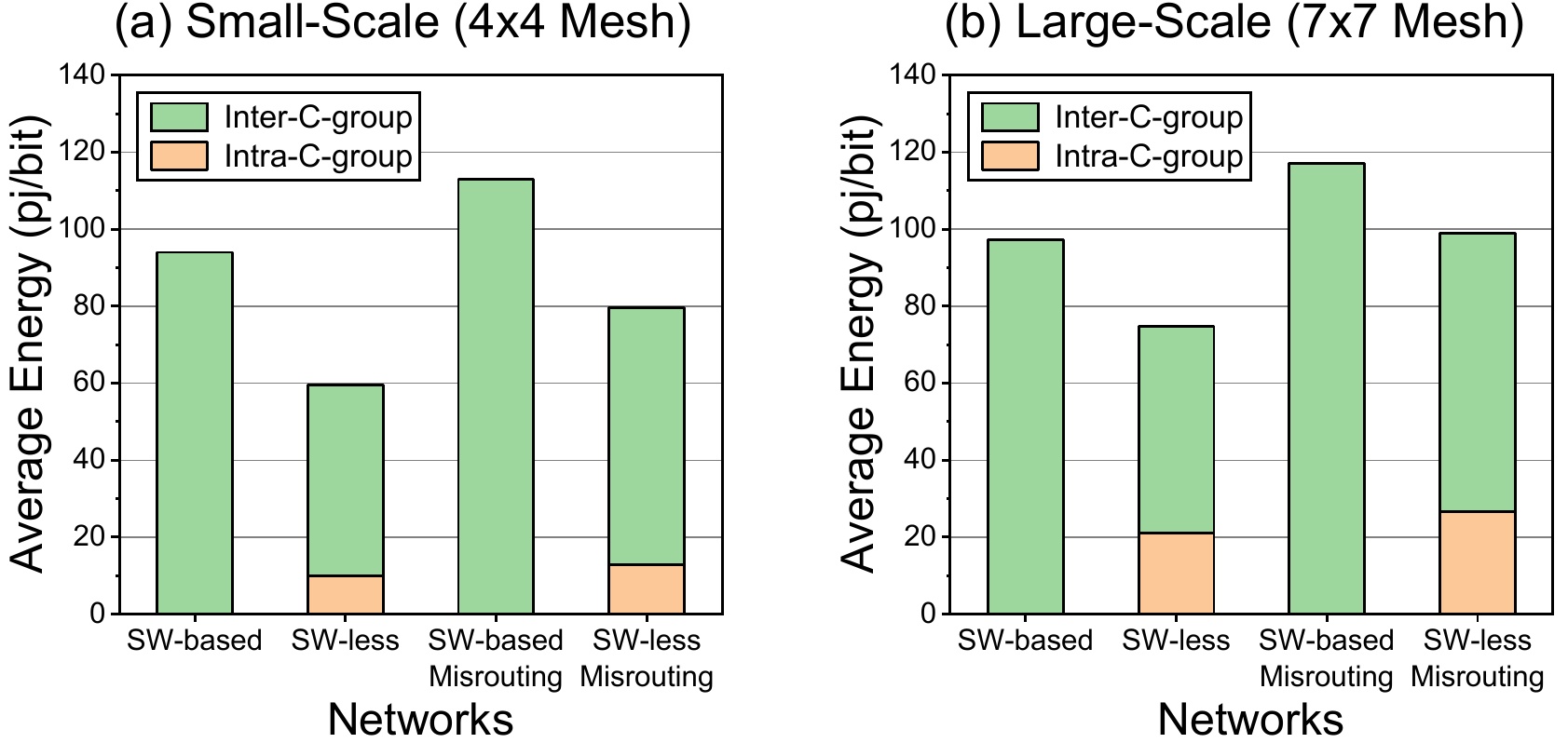}
  \caption{Average energy consumption per data transmission of minimal/non-minimal routing for small-scale and large-scale Dragonfly. \label{figure:energy}}
\end{figure}

\subsection{Power Consumption}
\label{sec:energy-evaluation}
Since the switching functionality is achieved by the intra-C-group network with numerous short-reach hops, it is not clear how the power consumption is affected. Considering modern switches have powerful software features, we evaluate the power consumption based on the energy per physical channel rather than directly comparing the chip power. As shown in TABLE~{\ref{tab:hop}}, the energy consumption of $H_l / H_g$, $H_{sr}$, and $H_\text{on-chip}$ is estimated at 20, 2, and 0.1 pj/bit, respectively. For simplicity, we assume an intra-C-group hop takes 1pj/bit on average. Uniform traffic is performed on topologies of different scales, and the trace of each packet is collected. As shown in Fig.~{\ref{figure:energy}}, the average energy consumption per data transmission is calculated based on the average hop count. For the small-scale Dragonfly, the energy overhead on the 4$\times$4 2D-mesh-on-wafer is small compared with the energy reduction from eliminating switches. For large-scale Dragonfly, since the diameter of 2D-mesh-on-wafer is larger, the energy overhead can be significant, especially for non-minimal routing. However, high-radix switches are also based on NoCs~\cite{DeSensi_InDepthAnalysisSlingshot_2020,Ahn_ScalableHighradixRouter_2013}, which also introduce extra energy consumption. In conclusion, eliminating switches can reduce the total energy consumption for both small/large-scale networks and minimal/non-minimal routing.

\section{Summary}
Wafer-scale integration provides high-density, low-latency, and high-bandwidth connectivity among tens of chips, thus promising to support direct high-radix networks without high-radix switches. In this paper, we propose a scalable wafer-based interconnection architecture for large-scale supercomputers. By utilizing distributed high-bandwidth networks-on-chip-on-wafer, costly high-radix switches of the Dragonfly topology are eliminated while increasing local throughput and maintaining global throughput. We also introduce baseline and improved deadlock-free minimal/non-minimal routing algorithms with only one additional virtual channel against traditional Dragonfly. Discussion and evaluations show that the switch-less Dragonfly is implementable, cost-effective, high-performance and scalable. The proposed wafer-based switch-less approach can be applied to other switch-based direct topologies and is promising to power future large-scale supercomputers.

\section{Acknowledgments}
This work is partially supported by the Wafer-Scale Silicon-Optic Interconnected System (2022YFB2804100) and the National Natural Science Foundation of China (20211710187).


\bibliographystyle{IEEEtran}
\bibliography{refs}


\end{document}